\documentclass{article}

\usepackage{PRIMEarxiv}

\usepackage[utf8]{inputenc} 
\usepackage[T1]{fontenc}    
\usepackage{hyperref}       
\usepackage{url}            
\usepackage{booktabs}       
\usepackage{amsfonts}       
\usepackage{nicefrac}       
\usepackage{microtype}      
\usepackage{lipsum}
\usepackage{fancyhdr}       
\usepackage{graphicx}       
\usepackage{natbib}

\graphicspath{{media/}}     

\title{A Deep-Learning-Based Framework for Focal Mechanism Determination 
and Its Application to the 2022 Luding Earthquake Sequence
}

\author{
  Ziye Yu \\
  Institute of Geophysics, China Earthquake Administration \\
  Beijing, China\\
  \texttt{yuziye@cea-igp.ac.cn,} \\
   \And
  Yuqi Cai \\
  Institute of Geophysics, China Earthquake Administration \\
  Beijing, China\\
  \texttt{caiyuqiming@foxmail.com} \\
}

\begin{document}
\maketitle

\begin{abstract}
P-wave first-motion polarity plays a critical role in resolving focal mechanisms of small to moderate earthquakes ($M \leq 4.5$). High-quality focal mechanism solutions for abundant small events can greatly enhance our understanding of regional tectonics, fault geometries, and stress-field characteristics. In this study, we develop an automated focal mechanism determination framework that integrates deep neural networks with P-wave first-motion polarity observations, and apply it to the 2022 Luding earthquake sequence. The model is trained on 12~years (2009--2020) of manually annotated 100~Hz waveform records from the China National Seismic Network, achieving a polarity recall of 97.4\% and a precision of 98.5\%. After automatically determining the first-motion polarities, we invert focal mechanisms using the HASH method. The resulting focal mechanism solutions show high consistency with mapped fault structures in the Luding region, demonstrating the reliability and applicability of the proposed workflow.
\end{abstract}


\section{Introduction}

Focal mechanism solutions provide essential information on fault strike, dip, and rake angles, forming the basis for understanding regional stress fields, crustal deformation, tectonic structures, and seismic hazard assessment \citep{Gephart1984Stress, Michael1984SlipStress, MartinezGarzon2016StressInversion, Xu1979StressBJ, Sheng2022FocalIdentification, Jia2012SouthwestParameters}. Small to moderate earthquakes ($M \leq 4.5$) occur in large numbers and exhibit dense spatiotemporal distributions; they are sensitive to small-scale fault structures and stress perturbations. As a result, these events play an important role in investigating crustal dynamics \citep{Yang2022LudingProcess} and assessing the likelihood of future large earthquakes \citep{Ming2003SmallEQMechanism, Zheng1994TaiwanMT, Cui1999SouthwestStress, Wan2011TaiwanStress}.

Traditional methods for determining focal mechanisms can be broadly categorized into two groups. The first group comprises waveform-inversion approaches, which estimate source parameters by fitting synthetic waveforms to observed records. These methods are widely used for moderate-to-large earthquakes \citep{Zhao1994BroadbandSource, Zhu1996SourceEstimation, Zhu2016LushanMTI} and can produce stable results even in regions with sparse station coverage. However, waveform inversion requires accurate velocity models and high signal-to-noise ratios, and its effectiveness is limited for small events because their radiated energy is concentrated at higher frequencies \citep{Tan2007ShortPeriod, Uchide2020JapanDeepLearning}. For small earthquakes, the rupture duration is short and the source dimension is small, such that the events can often be approximated by double-couple sources \citep{Lay1995ModernSeismology, Yang2022LudingProcess}. Because the waveforms of small earthquakes are generally simple, focal mechanism solutions obtained from P-wave first-motion polarities are often consistent with those derived from waveform fitting \citep{Li2023FocMechFlow, Shang2020BayesianSmallEQ}.

P-wave first-motion polarity has clear physical meaning, does not strongly depend on velocity-model accuracy, and can be inverted through grid search, making it a robust observable for determining focal mechanisms. Methods such as FPFIT \citep{Reasenberg1985FPFIT} invert focal mechanisms by fitting first-motion data mapped onto the focal sphere. However, uncertainties in velocity models and polarity picking introduce non-negligible errors \citep{Hardebeck2002HASH}. To address these limitations, the HASH method \citep{Hardebeck2002HASH, Hardebeck2003SPRatio} incorporates model perturbations and P/S amplitude ratios, resulting in more stable solutions. Previous studies have shown that well-distributed stations and reliable polarity readings substantially improve focal mechanism accuracy \citep{Tu2014XiningMechanism}. Conversely, for small near-source events, velocity-model uncertainty can greatly affect takeoff-angle calculations and degrade solution quality \citep{Yu2009GridSearch}. To mitigate uneven station coverage, probabilistic approaches and grid-based spatial parameterization have also been proposed \citep{Xu1995GridProbability}.

With the deployment of dense seismic networks, large volumes of high-quality waveform data have become available. However, manual annotation of first-motion polarities is labor-intensive, motivating the development of automated polarity classifiers. Traditional automatic approaches, such as zero-crossing counts \citep{Chen2016PhasePApy} and Bayesian algorithms \citep{Pugh2016BayesianPolarity}, rely on hand-crafted features and are sensitive to phase-picking accuracy. In contrast, data-driven deep learning methods have demonstrated exceptional performance in computer vision and natural language processing, and have been increasingly applied in seismology, including phase picking \citep{Yu2022LPPN, Yu2022Benchmark}, focal mechanism estimation \citep{Kuang2021DeepMechanism}, and earthquake classification \citep{Ku2020AttentionEQ}. Several studies have confirmed the effectiveness of neural networks for P-wave polarity determination \citep{Ross2018DeepPolarity, Chakraborty2022PolarCAP, Uchide2020JapanDeepLearning, Li2023FocMechFlow}.

To efficiently compute focal mechanisms for small to moderate earthquakes ($M \leq 4.5$) recorded by dense seismic arrays, we develop an automated focal mechanism determination framework based on P-wave first-motion polarity. We employ a recurrent-neural-network-based phase picker for accurate arrival-time estimation \citep{Yu2022LPPN}, and design a multi-layer residual network for polarity classification. The model is trained using 11~years (2009--2019) of manually labeled data from the China National Seismic Network (CSN) and evaluated on independent manually annotated data from 2020. Tests show that the trained model achieves performance comparable to manual classifications, and remains robust even when the P-wave pick contains timing errors. Finally, we invert focal mechanisms using the HASH grid-search algorithm \citep{Hardebeck2002HASH}. Based on the Luding earthquake sequence catalog of \citep{Liu2022LudingAftershocks}, we compute focal mechanisms and compare them with waveform-based solutions reported by \citep{Yang2022LudingProcess}.

\section{Methods and Data}

\subsection{Training Data}

\begin{figure}[h]
\centering
\includegraphics[width=\linewidth]{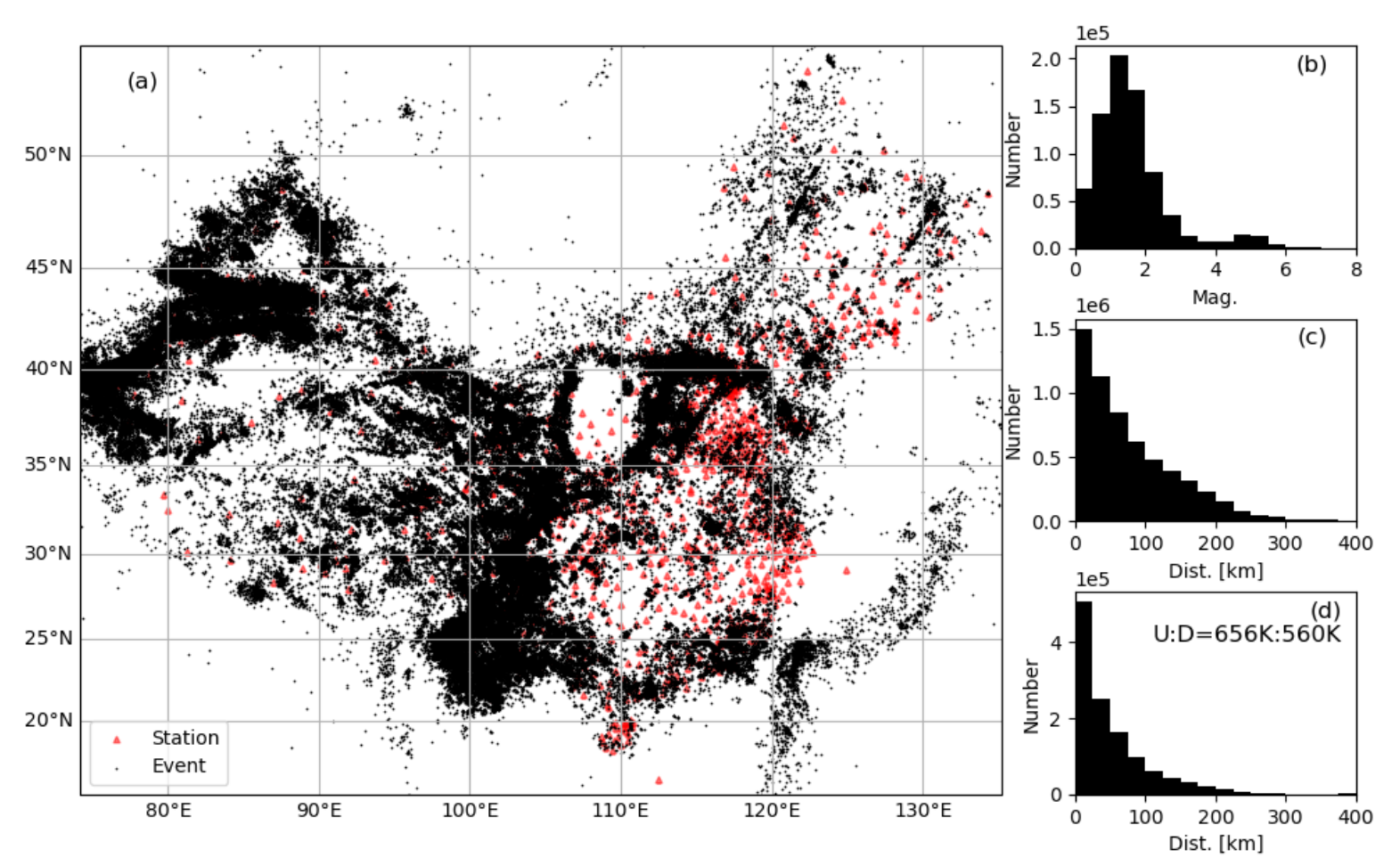}
\caption{\textbf{Data statistics for 2009--2020.}
(a) Distribution of stations (red triangles) and earthquakes (black dots);
(b) earthquake magnitude distribution;
(c) epicentral distance distribution of stations with Pg-phase picks;
(d) epicentral distance distribution of stations with polarity-labeled Pg phases.}
\label{fig:data_statistics}
\end{figure}

We use 11~years (2009--2020) of manually annotated data from the China Seismic Network (CSN) as the training dataset. All waveform records are vertical-component (Z) data sampled at 100~Hz. The dataset contains 1,350 fixed stations, 751,774 earthquake events, 5,908,236 Pg-phase picks, among which 1,217,021 Pg phases are labeled with first-motion polarity. The statistical distribution of the dataset is shown in Figure~\ref{fig:data_statistics}.

Figure~\ref{fig:data_statistics} illustrates:  
(a) the spatial distribution of stations (red triangles) and earthquakes (black dots);  
(b) magnitude distribution of events;  
(c) epicentral distance distribution for Pg phases;  
(d) epicentral distance distribution for polarity-labeled Pg phases.

As shown in Figure~\ref{fig:data_statistics}a, the combined distribution of earthquakes and stations ensures comprehensive coverage across mainland China, enabling the model to generalize well within this region. The proportions of upward and downward first-motion polarities are nearly balanced (approximately 1:1; Figure~\ref{fig:data_statistics}d), and therefore additional sampling re-balancing is unnecessary.

To evaluate the polarity-detection performance, we divide the dataset into training and testing subsets. Data from 2009--2019 serve as the training set, while data from 2020 serve as the independent test set. The test dataset includes 82,921 earthquake events, 702,010 P-wave picks, and 131,898 manually labeled first-motion polarities.

\subsection{Focal Mechanism Determination Framework}

The focal mechanism inversion framework consists of five major modules:

\begin{enumerate}
    \item \textbf{Data extraction:} For each event in the earthquake catalog, we compute P-wave travel times to all stations using TauP \citep{Crotwell1999TauP}, and extract 95~s of three-component waveform data from the continuous waveform archive.

    \item \textbf{Phase picking:} We apply the high-accuracy RNN-based phase picker trained on CSN data \citep{Yu2022Benchmark} to estimate precise P-wave arrival times from the extracted 95~s windows.

    \item \textbf{Preprocessing:} We extract 10.23~s of Z-component waveform data centered near the picked arrival and normalize it using its mean and standard deviation. During training, the P-arrival is randomly shifted within $\pm$2.56~s from the window center, enabling the polarity model to tolerate significant arrival-time errors.

    \item \textbf{First-motion polarity determination:} The preprocessed waveform is fed into the polarity-classification network, which outputs both the polarity (up/down) and its confidence. Signal-to-noise ratio (SNR), takeoff angle, and model confidence can be further used for polarity-quality assessment (see Section~Discussion).

    \item \textbf{Focal mechanism inversion:} Polarity results and station coordinates are provided to the HASH algorithm \citep{Hardebeck2002HASH}, which performs grid-search inversion to estimate focal mechanisms for small to moderate earthquakes.
\end{enumerate}

Figure~\ref{fig:workflow} presents the full automated workflow. Once the earthquake catalog is provided, all subsequent steps—including P-wave picking and polarity classification traditionally requiring extensive manual effort—are performed automatically by deep neural networks. P-wave arrival times are obtained using the RNN phase picker \citep{Yu2022Benchmark}, trained on the same 100~Hz CSN dataset described in Section~2.1. Because the polarity network tolerates P-arrival errors of up to $\pm2.56$~s, the travel-time predictions from TauP \citep{Crotwell1999TauP} can be used directly as substitutes for P-arrival picks if necessary.

\begin{figure}[h]
\centering
\includegraphics[width=0.4\linewidth]{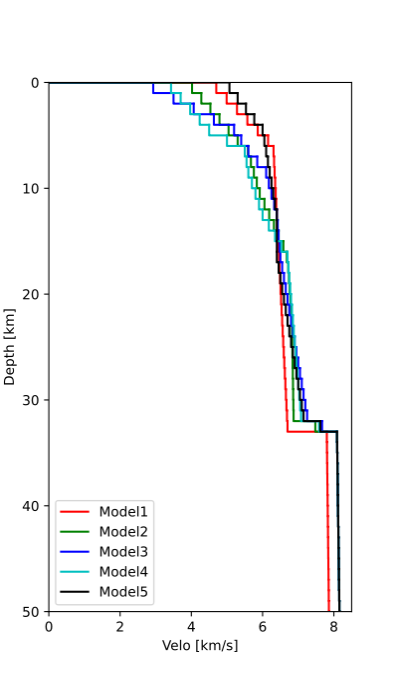}
\caption{\textbf{P-wave velocity models used in the grid-search process.}
The five P-wave velocity structures incorporated into the HASH inversion to enhance robustness of focal mechanism solutions.}
\label{fig:pwave_models}
\end{figure}

For focal mechanism inversion, we employ the HASH method \citep{Hardebeck2002HASH}, a robust grid-search algorithm that allows multiple velocity models to be incorporated to improve solution stability. We use five P-wave velocity models during grid search, shown in Figure~\ref{fig:pwave_models}.

\subsection{Polarity-Detection Model}

The P-wave picker and the polarity classifier are two independent models. The polarity model focuses solely on determining the first-motion direction. For each sample, we extract 1024 samples (10.23~s) of Z-component waveform centered around the P arrival. The preprocessing step follows standard practice: the waveform is normalized using its mean and standard deviation.

Polarity labels are encoded as integers (0 for upward, 1 for downward). Because the preceding P-wave picker may introduce timing errors, the training process randomly shifts the P arrival within a $\pm2.56$~s window. The polarity-detection model is a multi-layer residual convolutional network, and its architecture is summarized in Table~\ref{tab:arch}.

\begin{table}[h]
\centering
\caption{Architecture of the first-motion polarity classification model (base channel width $F=16$).}
\label{tab:arch}
\begin{tabular}{lcccc}
\hline
Layer & Input Channels & Output Channels & Stride & Activation \\
\hline
Conv & 1 & $F$ & 1 & ReLU \\
Conv+ResNet & $F$ & $2F$ & 2 & ReLU \\
Conv+ResNet & $2F$ & $3F$ & 2 & ReLU \\
Conv+ResNet & $3F$ & $4F$ & 2 & ReLU \\
Conv+ResNet & $4F$ & $5F$ & 2 & ReLU \\
Conv+ResNet & $5F$ & $6F$ & 2 & ReLU \\
Conv+ResNet & $6F$ & $7F$ & 2 & ReLU \\
Conv+ResNet & $7F$ & $8F$ & 2 & ReLU \\
Conv+ResNet & $8F$ & $9F$ & 2 & ReLU \\
Conv+ResNet & $9F$ & $10F$ & 2 & ReLU \\
Conv+ResNet & $10F$ & $22F$ & 2 & ReLU \\
Linear & -- & 2 & -- & Softmax \\
\hline
\end{tabular}
\end{table}

As shown in Table~\ref{tab:arch}, each block consists of a downsampling convolution followed by a residual module. Downsampling convolutions increase the number of feature channels, enabling richer representations, whereas residual modules increase network depth and improve accuracy. All convolutional layers are followed by batch normalization to accelerate training, and the kernel size is set to 5, which is effective for seismic waveform processing. The final fully connected layer outputs class probabilities for the two polarity types.

The model is trained using the cross-entropy loss function and optimized with the Adam algorithm \citep{Kingma2014Adam}. Training is performed on an NVIDIA A100 GPU for 200,000 iterations (approximately 24 hours, limited mainly by data-loading speed), with a batch size of 32.

\section{Results}

\subsection{Performance of the Polarity-Detection Network}

To evaluate the accuracy of the first-motion polarity detection network, we randomly select 20,000 manually labeled polarity samples from fixed stations of the China Seismic Network (CSN) in the test dataset. For quantitative assessment, we define four prediction categories: true positives (TP), false positives (FP), false negatives (FN), and true negatives (TN), where “positive’’ corresponds to upward first motion and “negative’’ to downward first motion.

\begin{figure}[htbp]
\centering
\includegraphics[width=0.4\linewidth]{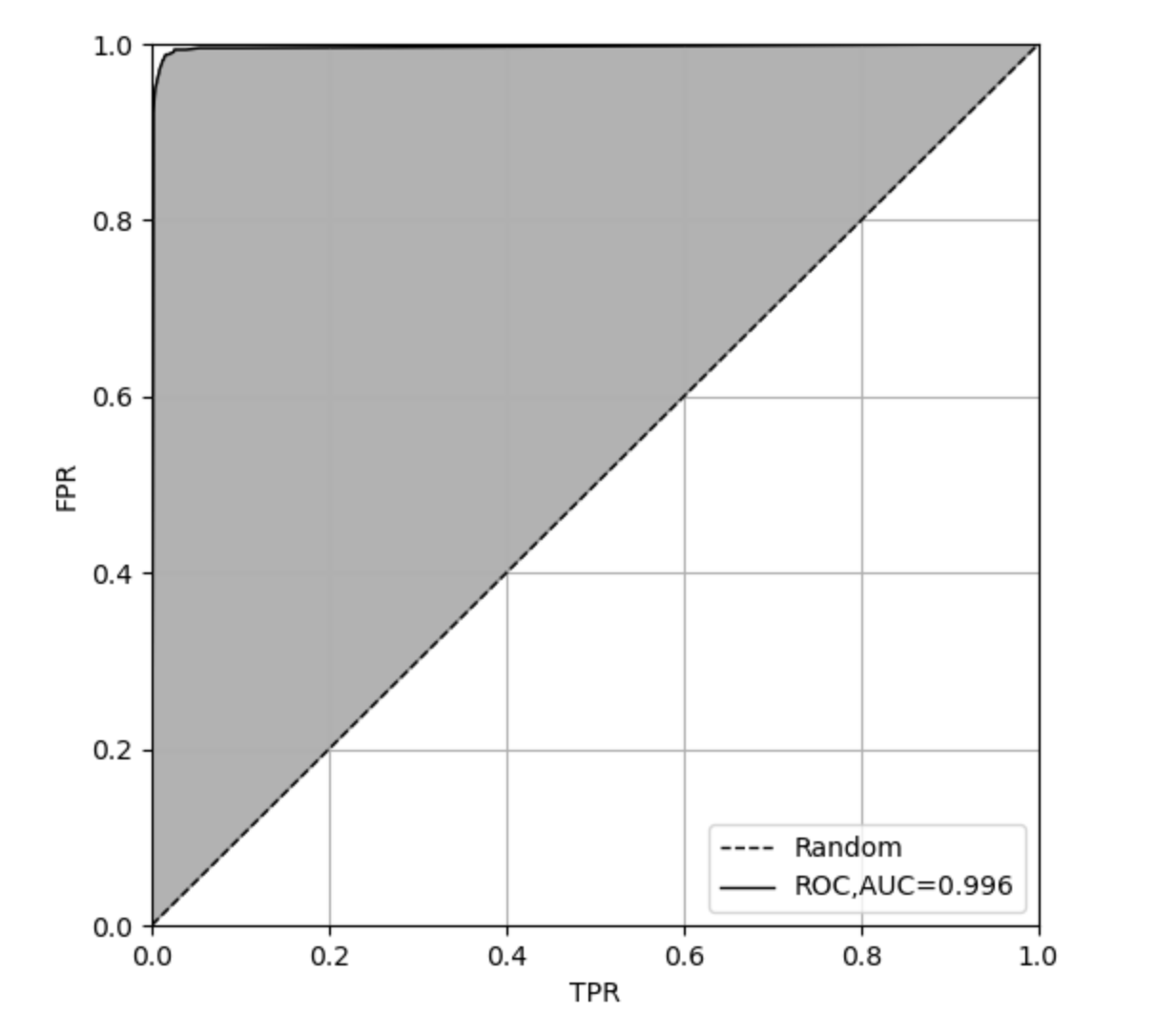}
\caption{\textbf{ROC curve for first-motion polarity detection.}
Receiver operating characteristic (ROC) curve of the polarity-classification model, illustrating performance across different probability thresholds.}
\label{fig:roc_curve}
\end{figure}

We center each P-wave phase in the 10.23~s detection window and compute precision, recall, and F1-score. The results are summarized in Table~\ref{tab:stat}. To examine the impact of different probability thresholds, we compute true-positive and false-positive rates across varying confidence levels and plot the receiver operating characteristic (ROC) curve shown in Figure~\ref{fig:roc_curve}.

\begin{table}[htbp]
\centering
\caption{Accuracy of the first-motion polarity detection model.}
\begin{tabular}{lc}
\hline
\label{tab:stat}

Metric & Value \\
\hline
Recall & 0.974 \\
Precision & 0.985 \\
F1-score & 0.980 \\
\hline
\end{tabular}
\end{table}

Figure~4 shows the ROC curve of the polarity-detection model. The accuracy exceeds 97\% and the AUC reaches 0.996, indicating excellent classification robustness across different probability thresholds. The model’s performance approaches the accuracy of manual picking.

To evaluate performance under arrival-time uncertainty, we test three arrival-time error scenarios:  
(1) 0~s,  
(2) $\pm0.1$~s,  
(3) $>1$~s.  

For each scenario, we further analyze results under different signal-to-noise ratios (SNR $\leq 30$~dB and SNR $> 30$~dB). We compare our results with those reported by FocMechFlow \citep{Li2023FocMechFlow}; SNR is computed following the same procedure. The comparison is shown in Table~\ref{tab:comp}.

\begin{table}[htbp]
\centering
\label{tab:comp}
\caption{Comparison of polarity-detection accuracy under different timing errors and SNR conditions.}
\begin{tabular}{lcccc}
\hline
Method & SNR & 0~s & $\pm$0.1~s & $>$1~s \\
\hline
Our model & All & 97.75\% & 98.06\% & 98.10\% \\
 & SNR$\le$30 dB & 96.46\% & 98.04\% & 97.91\% \\
 & SNR$>$30 dB & 98.76\% & 98.09\% & 98.46\% \\
\hline
Li et al. (2023) & All & 98.49\% & 84.26\% & -- \\
 & SNR$\le$30 dB & 97.57\% & 87.69\% & -- \\
 & SNR$>$30 dB & 99.60\% & 81.12\% & -- \\
\hline
\end{tabular}
\end{table}

The results show that our model performs comparably to FocMechFlow \citep{Li2023FocMechFlow} when arrival-time errors are small and SNR is high. However, when arrival-time errors exceed 1~s, our model outperforms theirs by more than 10\%, demonstrating robustness that allows the use of simple travel-time estimates such as TauP \citep{Crotwell1999TauP} as substitutes for precise P-wave arrival picks. 

\subsection{Application to the 2022 Luding Earthquake Sequence}

According to the China Earthquake Networks Center, an $M_\mathrm{s} 6.8$ earthquake occurred on 5 September 2022 near Luding County, Sichuan Province (102.08°E, 29.59°N), with a focal depth of 16~km. By 9 September 2022, a total of 6,211 earthquakes had been detected, with magnitudes ranging from ML$-0.5$ to ML$3.0$ (40 events with ML$\ge$3.0). We select 57 stations within 200~km of the mainshock, including 54 fixed stations of the Sichuan Digital Seismic Network and 3 temporary stations from the Chongqing and Yunnan networks \citep{Liu2022LudingAftershocks}. These stations provide a dense dataset suitable for focal mechanism inversion. 

\begin{figure}[h]
\centering
\includegraphics[width=0.5\linewidth]{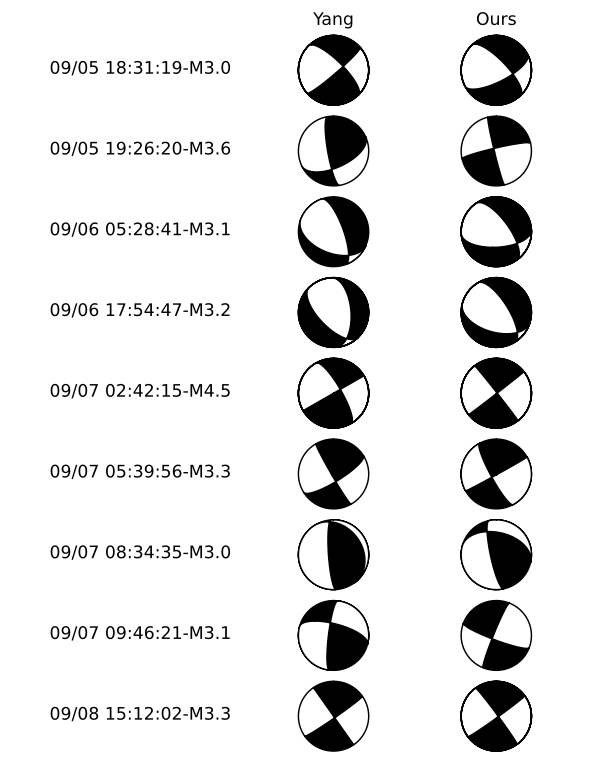}
\caption{\textbf{Comparison of focal mechanism solutions for the Luding earthquake sequence with previous work \citep{Yang2022LudingProcess}.}
}
\label{fig:luding_comparison}
\end{figure}

Focal mechanisms are inverted using the HASH algorithm \citep{Hardebeck2002HASH} with a grid interval of 1° and five different velocity models (Figure~\ref{fig:pwave_models}). The inversion results are shown in Figure~\ref{fig:luding_comparison}.

In total, 107 focal mechanisms are successfully determined, including the mainshock. Among them, 25 events have ML$>$3.0, and 82 events have ML$<$3.0, with the smallest magnitude being ML~0.4. This indicates that the proposed workflow can compute focal mechanisms for a large number of small to moderate earthquakes.

\begin{figure}[h]
\centering
\includegraphics[width=0.4\linewidth]{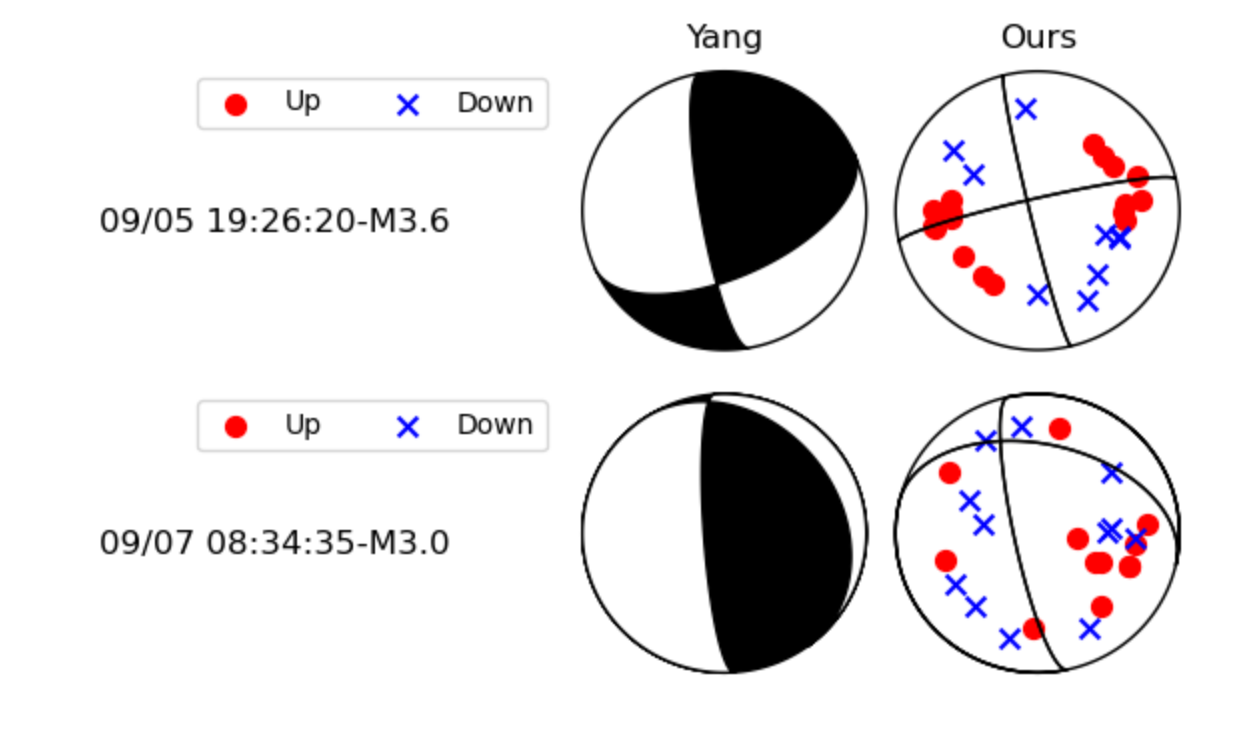}

\caption{\textbf{Events with large discrepancies from manually determined focal mechanisms.}
(a) Earthquake that occurred on 5 September 2022 at 19:26:20; 
(b) earthquake that occurred on 7 September 2022 at 08:34:35. 
These two events show the largest differences when compared with manual results, primarily due to low-quality first-motion waveforms.}
\label{fig:large_discrepancy_events}
\end{figure}

To validate the reliability of the inversion results, we compare nine events with focal mechanisms \citep{Yang2022LudingProcess}. The comparison is summarized in Figure~\ref{fig:luding_comparison}. Two events with relatively large discrepancies are further examined (Figure~\ref{fig:large_discrepancy_events}).

\begin{figure}[h]
\centering
\includegraphics[width=0.4\linewidth]{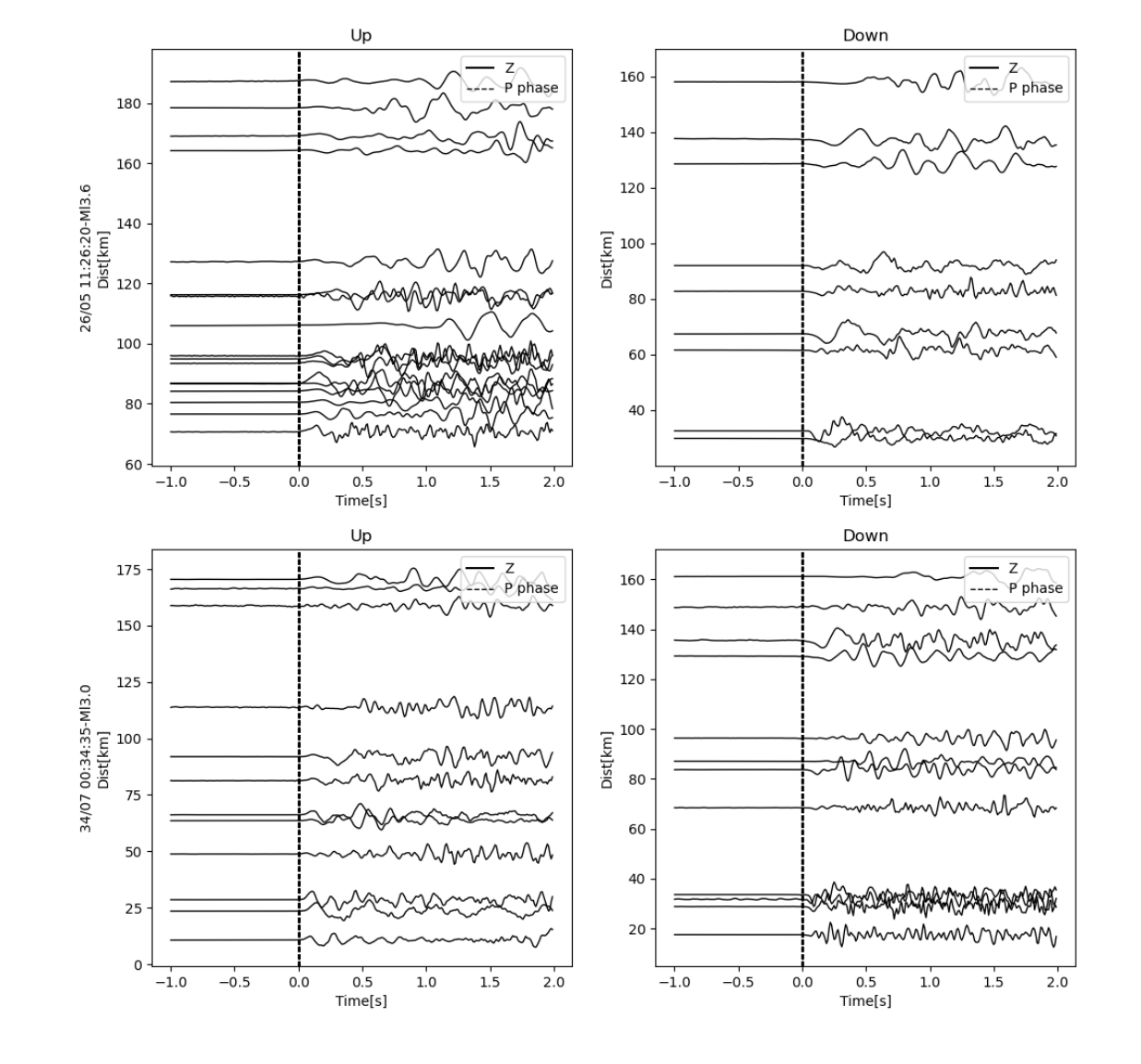}

\caption{\textbf{Waveforms used for events with large discrepancies from manual solutions.}
(a) Waveforms of the earthquake that occurred on 5 September 2022 at 19:26:20;
(b) waveforms of the earthquake that occurred on 7 September 2022 at 08:34:35.
In both cases, the P-wave onsets and first-motion polarities are difficult to determine, which can lead to mismatches between automated and manually derived focal mechanisms.}
\label{fig:waveform_large_discrepancy}
\end{figure}

For these events, the first-motion waveforms used in the inversion are shown in Figure~\ref{fig:waveform_large_discrepancy}. In the case with the largest discrepancies, P-wave first motions are of low quality, making manual or automatic determination difficult. Such events are typically excluded in traditional analyses.

The HASH inversion includes a solution-quality classification (A--D) based on the variance of acceptable solutions, where A represents the most stable solution and D represents the least constrained. For the event shown in Figure~\ref{fig:large_discrepancy_events}a, the HASH quality grade is B, suggesting that the discrepancy may result from manual-processing uncertainty in the reference solution. For the event in Figure~\ref{fig:large_discrepancy_events}b, station coverage is sparse west of the epicenter, leading to weak constraints; its HASH grade is D.

\begin{figure}[h]
\centering
\includegraphics[width=0.9\linewidth]{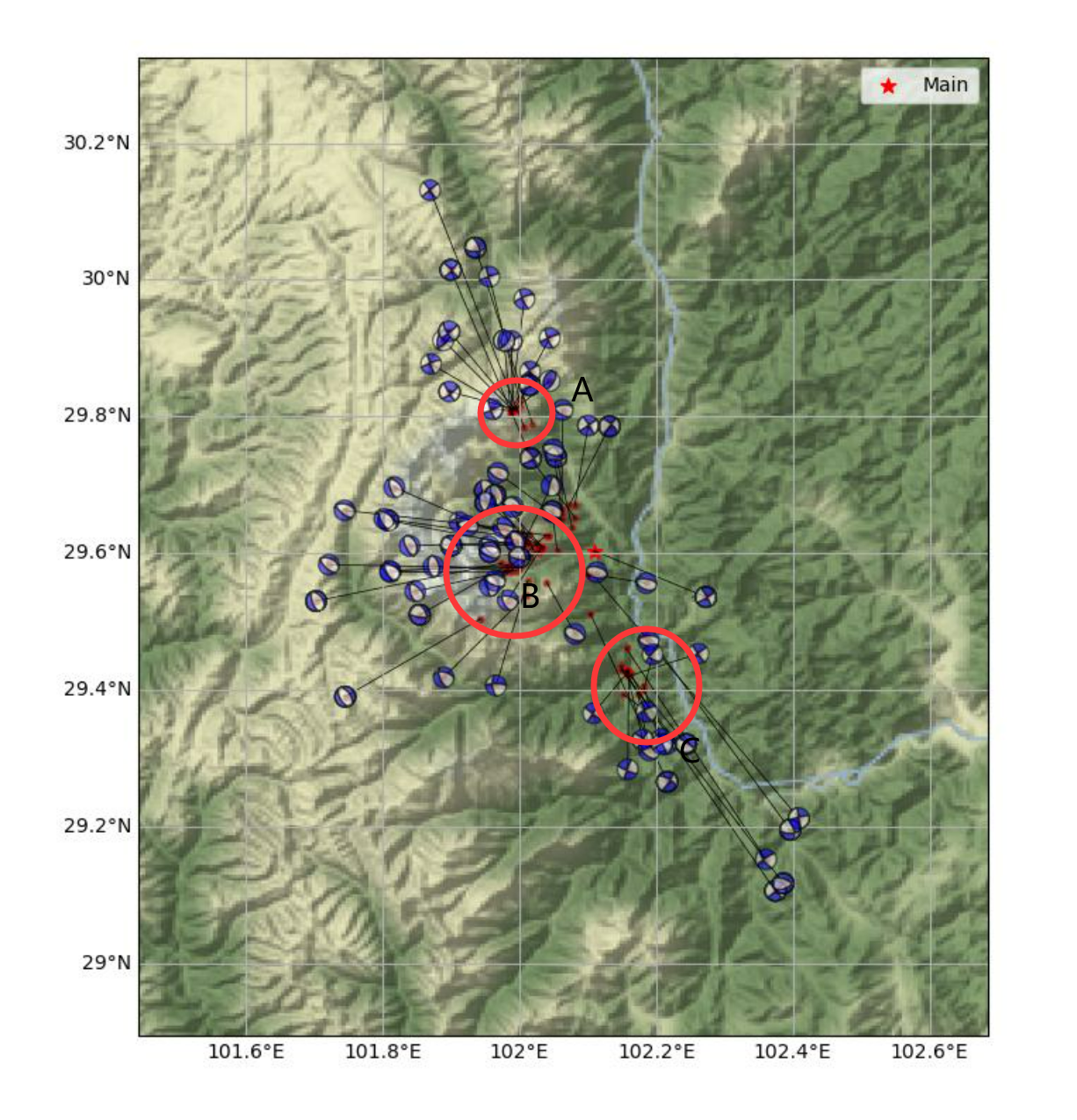}
\caption{\textbf{Spatial distribution of focal mechanisms for 107 earthquakes.}
Map view of the 107 inverted focal mechanisms in the 2022 Luding earthquake sequence. The aftershocks cluster into three main groups, with distinct dominant faulting styles in each cluster, highlighting the segmentation of fault structures in the study area.}
\label{fig:focal_mech_107}
\end{figure}

We plot the spatial distribution of the 107 inverted focal mechanisms in Figure~\ref{fig:focal_mech_107}. The aftershocks cluster into three distinct seismicity groups:  
Group~A (northwest of the mainshock) dominated by strike-slip mechanisms;  
Group~B (near the mainshock) dominated by normal faulting, differing from the mainshock mechanism;  
Group~C further east, again dominated by strike-slip mechanisms.

\section{Discussion}

In the focal mechanism determination framework developed in this study, the P-wave arrival time and first-motion polarity are computed by two independent deep-learning models. Both models are trained and evaluated using 12~years of manually annotated data from the national fixed-station network \citep{Yu2022Benchmark}, providing strong generalization capability across mainland China. By combining polarity estimates from multiple stations with the grid-search HASH algorithm \citep{Hardebeck2002HASH, Hardebeck2003SPRatio}, the workflow emulates the traditional manual procedure for polarity-based focal mechanism inversion, while replacing the non-automatable steps with deep neural networks. Integrated with seismic association algorithms, the framework enables real-time earthquake detection and parameter estimation. The successful application to the 2022 Luding earthquake sequence demonstrates that the designed automated workflow can be used for operational focal mechanism analysis.

We analyze the relationship among maximum nodal-plane separation angle, event magnitude, and solution-quality classification for all Luding focal mechanisms. The HASH solution-quality classification (A–D) is based on confidence, standard deviation of acceptable solutions, contradiction ratio, and station-distribution ratio \citep{Hardebeck2002HASH, Hardebeck2003SPRatio}. The criteria are summarized in Table~\ref{tab:quanti}.

\begin{table}[h]
\centering
\caption{Solution-quality classification criteria used in this study (interval intersections define each grade).}
\begin{tabular}{lcccc}
\label{tab:quanti}
Grade & Confidence & Std.~dev. (deg) & Contradiction ratio & Station-distribution ratio \\
\hline
A & (0.8, 1] & [0, 25) & [0, 0.15] & [0.5, 1] \\
B & (0.6, 0.8] & [25, 35] & (0.15, 0.2] & [0.4, 0.5) \\
C & (0.5, 0.6] & (25, 45] & (0.2, 0.3] & [0.3, 0.4) \\
D & Others & Others & Others & Others \\
\hline
\end{tabular}
\end{table}

As shown in Figure~10, when the number of available stations is small, the maximum separation angle tends to increase, and a larger proportion of events fall into the D-quality category. Improved station coverage would reduce the separation angle and increase the proportion of high-quality focal mechanism solutions.

In seismic cataloging workflows, the quality of manually assigned first-motion polarities is typically evaluated using two categories:  
(I) clear P-wave onset with reliable polarity;  
(E) unclear or ambiguous onset.  
Examples are shown in Figure~11. Category I waveforms exhibit clean first arrivals, whereas category E waveforms often lack a distinct onset, making polarity determination difficult even manually.

In this study, we evaluate three automated criteria for estimating first-motion quality:  
(1) a peak-amplitude angle metric,  
(2) the signal-to-noise ratio (SNR),  
(3) neural-network confidence.

For the peak-angle metric, we compare the angle associated with the first peak after the P arrival or a fixed-time window, whichever is smaller. For the SNR-based metric, we compute the ratio between the first post-arrival peak amplitude and the standard deviation of the one-second pre-arrival noise window, following common cataloging practice. The confidence-based metric directly uses the polarity-classifier output probability, reflecting the model-assessed certainty of each prediction.

To compare automatic and manual quality assessments, we analyze the distributions of peak angles, SNR, and model confidence for manually labeled I and E categories (Figure~12). While both peak-angle and confidence metrics show distinguishable trends between I and E events, the SNR metric provides the clearest separation.

Based on the statistical distributions, we evaluate the performance of each metric using fixed thresholds, summarized in Table~6.

\begin{table}[h]
\centering
\caption{Performance of three quality-assessment criteria for distinguishing manual I/E labels.}
\begin{tabular}{lccc}
\hline
Method & Precision & Recall & F1-score \\
\hline
Angle ($>$10°) & 0.92 & 0.511 & 0.657 \\
Confidence ($>$0.9) & 0.672 & 1.00 & 0.804 \\
SNR ($>$7) & 0.864 & 0.844 & 0.854 \\
\hline
\end{tabular}
\end{table}

The SNR-based criterion provides the most balanced recall and precision among the three methods. Therefore, in the automated workflow, SNR is adopted as the primary polarity-quality indicator. In this study, we classify an event as high-quality (I) if both SNR~$>$~7 and confidence~$>$~0.9; all other cases are labeled as E, corresponding to unclear first-motion polarity.

\section{Conclusion}

In this study, we trained a deep neural network for automatic P-wave first-motion polarity determination using a large-scale, manually annotated dataset from mainland China. Building on this model, we developed an automated focal mechanism inversion workflow based on first-motion polarity. Compared with other automated polarity-detection approaches, our method tolerates P-wave arrival-time errors of up to 2.56~s, meaning that high-precision arrival picking is not required. This significantly simplifies the processing pipeline, as travel-time predictions such as TauP can be used directly in operational settings.

Applying the workflow to the 2022 Luding earthquake sequence, we successfully determined focal mechanisms for 174 aftershocks. The results are generally consistent with previous studies, and the workflow enables focal mechanism inversion for events as small as ML~1.0. This increases the number of usable focal-mechanism solutions and enhances the ability to characterize fault geometry and regional stress fields with greater detail.

The Luding aftershock sequence exhibits a typical mainshock–aftershock pattern. Spatially, the aftershocks cluster into three distinct groups, likely associated with three separate fault segments. Among these, cluster~B, located near the mainshock, shows relatively strong activity. The improved ability to derive focal mechanisms for small earthquakes provides valuable constraints for understanding fault segmentation and stress evolution in this region.

\section*{Open Research}

All data and code used in this study are openly available. The scripts for 
P-wave first-motion polarity classification, focal-mechanism inversion, and 
the full processing workflow are released under an open-source license at:

\begin{center}
\url{https://github.com/cangyeone/seismological-ai-tools/tree/main/p-polarity-and-focal-mac}
\end{center}

This repository contains preprocessing scripts (waveform extraction, P-wave
window selection), polarity-classification models, focal-mechanism inversion 
tools (grid-search and probabilistic implementations), configuration files, 
and all reproducible experiment scripts used in the manuscript.

All waveform data used in the demonstrations come from publicly available 
seismic networks and are cited in the Data and Resources section of this 
paper. All derived products (polarity labels, focal-mechanism solutions, 
and figures) can be fully reproduced using the scripts provided in the 
repository.

\bibliographystyle{unsrt}  
\bibliography{references}

\end{document}